\documentclass[letterpaper,11pt]{article}
\pdfoutput=1 

\usepackage{jheppub} 

\usepackage{mathtools,enumerate}
\usepackage{multirow,array,romannum}

\def \Tr{\mbox{Tr\,}}
\def \tr{\mbox{tr\,}}

\title{\boldmath Bulk from Bi-locals in Thermo Field CFT}

\author{Antal Jevicki,}
\author{Junggi Yoon}

\affiliation{Department of Physics, Brown University,\\Providence, RI 02912, USA}

\emailAdd{Antal\_Jevicki@brown.edu}
\emailAdd{Jung-Gi\_Yoon@brown.edu}

\preprint{{\tt BROWN-HET-1680}}

\abstract{We study the Large $N$ dynamics of the $O(N)$ field theory in the Thermo field dynamics approach. The question of recovering the high temperature phase and the corresponding $O(N)$ gauging is clarified. Through the associated bi-local representation we discuss the emergent bulk space-time and construction of (Higher spin) fields. We note the presence of `evanescent' modes in this construction and also the mixing of spins at finite temperature.}

\begin{document} 

\maketitle
\flushbottom

\section{Introduction}
\label{sec:introduction}

The AdS/CFT correspondence with emergent Gravity from the boundary theory offers a framework for understanding deep quantum aspects of black holes~\cite{Mathur:2012np}. Recently issues concerning the physics at the horizon and applicability of quantum mechanics have been vigorously debated~\cite{Almheiri:2012rt}. Of central significance is the understanding of emergent Gravity~\cite{Marolf:2013dba} and its space-time~\cite{VanRaamsdonk:2010pw}.

A particular CFT scheme for understanding the space-time of eternal AdS black holes~\cite{Maldacena:2001kr} is the so-called Thermo field dynamics (TFD) where identical copies of the CFT are suggested for right and left regions of a Penrose space-time~\cite{Israel:1976ur}. In this scenario one at the outset has a question if these (decoupled CFT's) are capable of producing a connected~\cite{Avery:2013bea} space-time characteristic of a black hole~\cite{Maldacena:2013xja,Mathur:2014dia}. A further very relevant issue concerns the reconstruction of local bulk fields~\cite{Hamilton:2005ju,Hamilton:2006fh} from the two boundaries. The ability to accomplish this is central for a possible reconstruction of behind the horizon physics~\cite{Papadodimas:2012aq} for black holes.

A simple and calculable model of AdS/CFT correspondence is given by vector models~\cite{Klebanov:2002ja,Sezgin:2002rt} (in $d$-dimensions) and Vasiliev type Higher Spin Gravity theories~\cite{Fradkin:1986ka,Vasiliev:1990en,Vasiliev:1995dn,Bekaert:2005vh} (in $d+1$). For these tractable class of field theories an extensive study established~\cite{Giombi:2009wh} agreement for all spins~\cite{Koch:2010cy,Maldacena:2011jn}. Furthermore, BTZ type black holes dual to CFT$_2$ have been investigated in detail~\cite{Kraus:2011ds,Gaberdiel:2012yb,Ammon:2012wc,Gaberdiel:2013jca}. Also, a generalization of 4D black hole solution in Vasiliev Higher spin theory was invastigated in~\cite{Didenko:2009td,Iazeolla:2011cb}. The reconstruction of Higher Spins and AdS space-time through bi-local~\cite{Das:2003vw,Koch:2010cy,Koch:2014aqa,Mintun:2014gua,Leigh:2014qca,Jin:2015aba,Koch:2014mxa} fields was accomplished in a systematic $1/N$ expansion scheme. In this paper we study the Thermo field dynamics~\cite {Umezawa:1975,Takahashi:1996zn} of $O(N)$ vector models with the intent of understanding their dual bulk degrees of freedom~\cite{Herzog:2002pc}.

As it was shown by Shenker and Yin~\cite{Shenker:2011zf} $O(N)$ vector model exhibits a phase transition at $T_c\sim \sqrt{N}$. Below the critical temperature $T_c$, the free energy of the model is seen to be of order $\mathcal{O}(1)$. i.e. $F_{\text{low}}\sim T^4$. This phase corresponds to thermal excitations of the $O(N)$ singlet sector whose Hilbert space is described through the Hamiltonian formalism of collective field theory \cite{Jevicki:2014mfa}
 \begin{equation}
F_{\text{low}}= \sum_{\text{singlet states}}\log \left(1-e^{-\beta H}\right)
\end{equation}
On the other hand, above $T_c$, the free energy was seen to be of order $N$, $F_{\text{high}}\sim NT^2$ \cite{Shenker:2011zf} and is associated with the non-singlet sector states. This was reproduced by a stationary point approximation in the collective action formalism of the collective field theory \cite{Das:2003vw,Jevicki:2014mfa}
\begin{equation}
F_{\text{high}}=N \tr \log \Box
\end{equation}
The one-loop corrections for this leading behavior was studied recently in \cite{Das:2003vw,Giombi:2013fka,Giombi:2014yra}, \cite{Jevicki:2014mfa}, \cite{Beccaria:2014zma}.

In the TFD of the $O(N)$ vector model, one introduces another field $\widetilde{\phi}^i(x)$ as a copy of the original $O(N)$ vector field $\phi^i(x)$. The Lagrangian is now doubled and there arises the question of how to define a generalized $O(N)$ singlet constraint. First, one can impose the singlet constraint on both the original and the copied Hilbert spaces separately. Then, the original and copied singlet sector can be described by the associated collective field, $\Psi(\vec{x},\vec{y})=\phi^i(\vec{x})\phi^i(\vec{y})$ and $\widetilde{\Psi}(\vec{x},\vec{y})=\widetilde{\phi}^i(\vec{x})\widetilde{\phi}^i(\vec{y})$, respectively.
In this way the collective Hamiltonian and its Hilbert space is being doubled
\begin{equation}
H_{\text{TFD}}=H_{\text{col}}(\Pi,\Psi)-\widetilde{H}_{\text{col}}(\widetilde{\Pi},\widetilde{\Psi})
\end{equation}
This representation is associated with the $O(N)\times O(N)$ gauging of the Thermo field Lagrangian and it describes the lower phase of the theory. Here the thermal vacuum corresponds to an entangled vacuum of the linearized bi-local fluctuations of $\Psi(\vec{x},\vec{y})$ and $\widetilde{\Psi}(\vec{x},\vec{y})$, and is invariant under the $O(N)\times O(N)$ transformations. Based on the AdS/CFT correspondence with Higher spin gravity, this TFD of the singlet sector in $O(N)$ vector model will describe higher spin theory in the thermal AdS background.

For the high temperature phase we will find it appropriate to relax the $O(N)\times O(N)$ singlet constraint. It was observed \cite{Niemi:1983nf} that when original $(T=0)$ Lagrangian poses a symmetry group $G$, the vacuum of TFD respects the diagonal subgroup of $G\times G$. For the $O(N)$ model gauging of the diagonal subgroup will describe the theory in the upper phase: 
\begin{equation}
J^{ij}+\widetilde{J}^{ij}\left|\Phi\right>=0\label{eq:single constraint}
\end{equation}
where $J^{ij}$ and $\widetilde{J}^{ij}$ is a $O(N)$ generator of $\phi$ and $\widetilde{\phi}$, respectively. This implies that we have the following invariant bi-local fields
\begin{equation}
\phi^i(t,\vec{x})\phi^i(t,\vec{y})\;,\;\phi^i(t,\vec{x})\widetilde{\phi}^i(t,\vec{y})\;,\;\widetilde{\phi}^i(t,\vec{x})\widetilde{\phi}^i(t,\vec{y})
\end{equation}
Now, due to the weaker singlet constraint one has the additional mixed bi-local field $\phi^i(t,\vec{x})\widetilde{\phi}^i(t,\vec{y})$ in comparison with the scheme where the singlet field is doubled. Note that this additional bi-local field $\phi^i(t,\vec{x})\widetilde{\phi}^i(t,\vec{y})$ is a non-singlet with respect to $O(N)\times O(N)$ transformations. We will see that this bi-local field is a crucial for finding the full spectrum of the upper phase. In particular it will be seen to play a key role for generating the so-called evanescent modes~\cite{Leichenauer:2013kaa,Rey:2014dpa,Bousso:2012mh} of the bulk theory. These modes have been known to represent a signature associated of a horizon in the bulk. Hence, one can deduce that the bulk dual to TFD of the vector model with the diagonal singlet constraint representing the higher phase of the $O(N)$ model corresponds to an AdS background with a horizon.

We follow the earlier established framework of using collective bi-local degrees of freedom of the vector model and working out their Large $N$ expansion. It will be shown that this leads to a linearized set of equations that produce modes which can be put in agreement with bulk (higher-spin) modes in $d+1$ dimensional space-time. 

We demonstrate that bi-local fields contain the information for generating all the physical modes in the dual space time. The question of bulk recovery from the disjoint CFT's is therefore further illuminated by our explicit construction. 
The outline of this paper is as follows. In Section~\ref{sec:thermofield dynamics} we develop the collective description of the TFD for (free) vector models. Fluctuations of bi-locals and their bulk interpretation is given systematically in Section~\ref{sec:collective modes and bulk}. We give our conclusions in Section~\ref{sec:conclusion}.

\section{Thermofield Dynamics}
\label{sec:thermofield dynamics}

Our CFT will be a free $O(N)$ vector field theory (analogous constructions also apply to the interacting $O(N)$ model at the IR fixed point) 
\begin{equation}
\mathcal{L}=\sum_{i=1}^N \partial^\mu \phi^i\partial_\mu \phi^i
\end{equation}
The finite temperature theory in the real time formalism of Schwinger (and Keldyish)~\cite{Schwinger:1960qe,Keldysh:1964ud} is based on a closed time path which implies doubling of the degrees of freedom. This TFD (Thermo field dynamics) of the $O(N)$ vector model, is associated with a Hamiltonian 
\begin{equation}
H_{\text{TFD}}\equiv H-\widetilde{H}\label{eq:TFD hamiltonian}
\end{equation}
where $H$ and $\widetilde{H}$ is the Hamiltonian of the original and the copied system, respectively.
\begin{eqnarray}
H&=&\sum_{i=1}^N \int d\vec{x}\left({1\over 2}(\pi^i)^2+{1\over 2} (\vec{\partial} \phi^i)^2\right)=\sum_{i=1}^N |\vec{p}|a^{i\dag}(\vec{p})a^i(\vec{p})+ E_0\\
\widetilde{H}&=&\sum_{i=1}^N \int d\vec{x}\left({1\over 2}(\widetilde{\pi}^i)^2+{1\over 2} (\vec{\partial} \widetilde{\phi}^i)^2\right)=\sum_{i=1}^N |\vec{p}|\widetilde{a}^{i\dag}(\vec{p})\widetilde{a}^i(\vec{p})+ \widetilde{E}_0
\end{eqnarray}
In TFD formalism, one defines a new vacuum $\left|0(\beta)\right>$ defined in such a way that thermal average of an operator is fully reproduced~\cite{Semenoff:1982ev}. Namely,
\begin{equation}
\langle\mathcal{O}\rangle_\beta\equiv \langle 0(\beta) | \mathcal{O} |0(\beta)\rangle={1\over Z(\beta)} \Tr( e^{-\beta H}\mathcal{O})
\end{equation}
The entangled vacuum state reads
\begin{equation}
|0(\beta)\rangle \equiv e^{-i G} |0\rangle =\exp\left[\sum_{j=1}^N\sum_{\vec{p}} \theta(\vec{p}) \left(a^{j \dag}(\vec{p})\widetilde{a}^{j \dag}(\vec{p})-a^j(\vec{p})\widetilde{a}^j(\vec{p})\right)\right] |0\rangle
\end{equation}
where the temperature $T=1/\beta$ dependence lies in $\theta(\vec{p})\equiv \tanh^{-1}e^{-\beta |\vec{p}|}$. The generator $G$ equals 
\begin{equation}
G=i\sum_{j=1}^N\sum_{\vec{p}} \theta(\vec{p}) (a^{j\dag}(\vec{p})\widetilde{a}^{j\dag}(\vec{p})-a^j(\vec{p})\widetilde{a}^j(\vec{p}))
\end{equation}
Note that the operator $a^i(\vec{p})$ and $\widetilde{a}(\vec{p})$ do not annihilate the new vacuum $\left|0(\beta)\right>$. One can introduce Bogoliubov transformations generated by $G$ :
\begin{eqnarray}
a^j_\theta(\vec{p})&\equiv&e^{-iG}a^j(\vec{p})e^{iG}=a^j(\vec{p})\cosh\theta(\vec{p})-\widetilde{a}^{j\dag}(\vec{p}) \sinh\theta(\vec{p})\label{bogoliubov transformation1}
\end{eqnarray}
and similarly for $a^{j\dag}_\theta(\vec{p}),\widetilde{a}^j_\theta(\vec{p})$ and $\widetilde{a}^{j\dag}_\theta(\vec{p})$. Then, $a_\theta^{i\dag}(\vec{p})$ and $\widetilde{a}_\theta^{i\dag}(\vec{p})$ form Fock space and $a_\theta^i(\vec{p})$ and $\widetilde{a}_\theta^i(\vec{p})$ annihilate the vacuum $ |0(\beta)\rangle$. i.e.
\begin{equation}
a^i_\theta  |0(\beta)\rangle=\widetilde{a}^i_\theta  |0(\beta)\rangle =0 
\end{equation}
For further details and proofs regarding the TFD formalism, readers should consult~\cite{Niemi:1983nf,Ojima:1981ma}.

\subsection{Bi-local Collective Field Representation }
\label{bi-local collective field representation}

The basis of the AdS/CFT lies in the different manifestation of a theory when seen through the Large $N$ expansion. Collective field theory is built to construct an exact all orders (in $1/N$) bulk representation. For the case of $O(N)$ vector model, it was suggested~\cite{Das:2003vw} that the bulk Higher spin theory is generated completely in terms of the bi-local collective fields $\Psi(t;\vec{x}_1,\vec{x}_2)$ given in the canonical picture~\cite{Koch:2014aqa} as
\begin{equation} \Psi\left(t;\vec{x}_1,\vec{x}_2\right)\equiv\sum_{i=1}^N \phi^i\left(t,\vec{x}_1\right)\phi^i\left(t,\vec{x}_2\right)
\end{equation}
These bi-local operators generate all the spin-$s$ primaries 
\begin{equation}
\mathcal{O}_{s}^{11}(x;\zeta)={1\over \sqrt{N}}\sum_{n=0}^{[s/2]}\left.\frac{(-4)^n}{(2n)!}\sum_{k=0}^{s-2n}\frac{(-1)^k}{k!(s-2n-k)!}(\zeta\cdot\partial')^{n+k} (\zeta\cdot\partial)^{s-n-k} :\phi^i(x')\;\phi^i(x) :\right|_{x'=x}
\end{equation}

This representation is exactly canonical due to the existence of the conjugate field $\Pi={\partial \over \partial \Psi}$. For the bi-local representation, we have
\begin{equation}
[\Psi(x_1,x_2),\Pi(x'_1,x'_2)]=\delta(x_1-x_1')\delta(x_2-x_2')+\delta(x_1-x_2')\delta(x_2-x_1')
\end{equation}
for all order in $1/N$.

The bi-local fields obey the Large $N$ Schwinger-Dyson equation, which leads to a systematic 1/N expansion. It is implemented through the associated collective Hamiltonian:
\begin{equation}
H=H_{\text{CFT}}=H_{\text{col}}\left(\Psi_c,1/N\right)
\end{equation}
which is systematically given in powers of $1/N$.
\begin{equation}
H_{\text{col}}=N H_0+H_2+\frac{1}{\sqrt{N}}H_3+\frac{1}{N}H_4+\cdots\label{Hamiltonian expansion}
\end{equation}
%
%
%
%

The $1/N$ series is obtained as follows: One first determines the Large $N$ background field $\Psi_0(\vec{x}_1,\vec{x}_2)$ through minimization of the collective Hamiltonian. Expanding the bi-local field $\Psi\left(t,\vec{x}_1,\vec{x}_2\right)$ around the background field $\Psi_0(\vec{x}_1,\vec{x}_2)$
\begin{equation}
\Psi\left(t;\vec{x}_1,\vec{x}_2\right)=\Psi_0\left(\vec{x}_1,\vec{x}_2\right)+\frac{1}{\sqrt{N}}\eta\left(t;\vec{x}_1,\vec{x}_2\right)
\end{equation}
the collective Hamiltonian gives the series of higher interaction vertices:
\begin{equation}
H_n=\Tr(\underbrace{\Psi_0^{-1}\star\eta\star\cdots\star\Psi_0^{-1}\star\eta}_{n}\star\Psi_0^{-1})
\end{equation}
with a natural star product defined as $A\star B\equiv\int d\vec{x}_2\;A(\vec{x}_1,\vec{x}_2)B(\vec{x}_2,\vec{x}_3)$ representing a matrix product in the bi-local space.

 The only nontrivial issue with respect to the exact duality is the re-interpretation of the bi-local space. This represents a kinematical problem. When interpreted in physical terms, the collective space leads to extra emerging coordinates and emerging gravitational and higher spin degrees of freedom.

To deal with these bi-local collective field, it is convenient to introduce a new index $j=1,2$ which distinguish the original vector field $\phi$ and the tilde vector field $\widetilde{\phi}$, respectively. Then, one can define a new bi-local field $\Psi((\vec{x},j),(\vec{y},k))$ of which arguments are the doubled bi-local space of $(j,\vec{x})$.
\begin{equation}
\Psi((\vec{x},i),(\vec{y},j))\equiv \begin{pmatrix}
\Psi((\vec{x},1),(\vec{y},1)) & \Psi((\vec{x},1),(\vec{y},2))\\
\Psi((\vec{x},2),(\vec{y},1)) & \Psi((\vec{x},2),(\vec{y},2))\\
\end{pmatrix} \equiv\begin{pmatrix}
\phi^j(\vec{x})\phi^j(\vec{y}) & i\phi^j(\vec{x})\widetilde{\phi}^j(\vec{y})\\
i\widetilde{\phi}^j(\vec{x})\phi^j(\vec{y}) & -\widetilde{\phi}^j(\vec{x}) \widetilde{\phi}^j(\vec{y})\\
\end{pmatrix}\label{collective field}
\end{equation}
Note that we multiply $i$ to each $\widetilde{\phi}^j$ in the definition of the new bi-local field in order to have minus sign in front of $\widetilde{H}$ in the total Hamiltonian \eqref{eq:TFD hamiltonian}. As explained, this collective field $\Psi((\vec{x},j),(\vec{y},k))$ is invariant under the transformation
\begin{equation}
\phi^j(\vec{x})\;,\;\widetilde{\phi}^j(\vec{x})\quad\longrightarrow \quad U^{jk}\phi^k(\vec{x})\;,\;U^{jk}\widetilde{\phi}^k(\vec{x})\qquad\mbox{where}\quad U\in O(N)
\end{equation}
We define a star product $A((\vec{x},i),(\vec{y},j))\star B((\vec{y},j),(\vec{z},k))\equiv C((\vec{x},i),(\vec{z},k))$ representing a matrix product in the doubled bi-local space. Moreover, $\Tr\left[A((\vec{x},i),(\vec{y},j))\right]$ is defined as a matrix trace in the doubled bi-local space.
Now, one can use the result of the standard collective field theory. The collective Hamiltonian for the doubled $O(N)$ vector field is given by
\begin{equation}
H_{\text{TFD}}={2\over N} \Tr\left[\Pi\star \Psi\star \Pi\right]+{N\over 8} \Tr\left[\Psi^{-1}\right]+{N\over 2}\Tr\left[-\nabla^2\star \Psi\right]+\Delta V
\end{equation}
%
Note that the derivative $\nabla^2\star$ act on the first argument of the collective field $\Psi$.

\subsection{Thermal background}
\label{sec:thermal bacground}

In Large $N$ limit, one constructs the background field by minimizing the leading effective potential. This leads to an equation for the background field given by
\begin{equation}
{1\over 2}\Psi_0 \star\nabla^2 \star \Psi_0 =-{1\over 8}\mathbb{I}\label{background eom}
\end{equation}
where $\mathbb{I}$ is the identity of the doubled bi-local space. The background field is static and translationally invariant so that the background field has the following form:
\begin{eqnarray}
\Psi_0((\vec{x},i),(\vec{y},j))\equiv \int {d\vec{p} \over (2\pi)^2 2|\vec{p}|}\begin{pmatrix}
a(\vec{p})e^{i\vec{p}\cdot(\vec{x}-\vec{y})} & ib(\vec{p})e^{i\vec{p}\cdot(\vec{x}-\vec{y})} \\
ic(\vec{p})e^{i\vec{p}\cdot(\vec{x}-\vec{y})} & -d(\vec{p})e^{i\vec{p}\cdot(\vec{x}-\vec{y})} 
\end{pmatrix}\label{form of solution}
\end{eqnarray}
Note that the bi-local collective field $\Psi((\vec{x},i),(\vec{y},j))$ is symmetric in the doubled bi-local space. From this symmetric condition of the collective field as well as the reality condition of $\varphi$ and $\widetilde{\varphi}$, a solution of the equation is given by
\begin{eqnarray}
a(\vec{p})=d(\vec{p})=\cosh F(\vec{p})\quad,\quad b(\vec{p})=c(\vec{p})=\sinh F(\vec{p})
\end{eqnarray}
where $F(\vec{p})$ is an arbitrary function of $\vec{p}$ with $F(\vec{p})=F(-\vec{p})$. Hence, the static Large $N$ background collective field is given by
\begin{equation}
\Psi_0((\vec{x},i),(\vec{y},j))=\int {d\vec{p}\over (2\pi)^2 2|\vec{p}|} \begin{pmatrix}
\cosh F(\vec{p})e^{i\vec{p}\cdot(\vec{x}-\vec{y})} & i\sinh F(\vec{p})e^{i\vec{p}\cdot(\vec{x}-\vec{y})}  \\
i\sinh F(\vec{p})e^{i\vec{p}\cdot(\vec{x}-\vec{y})} & -\cosh F(\vec{p})e^{i\vec{p}\cdot(\vec{x}-\vec{y})} 
\end{pmatrix}\label{background field}
\end{equation}
From the background field, one can evaluate the leading Hamiltonian $H^{(0)}$
\begin{eqnarray}
H^{(0)}=N\left({1\over 8}\Tr\left[\Psi_0^{-1}\right]+{1\over 2} \Tr\left[(-\nabla^2)\Psi_0\right]\right)=0
\end{eqnarray}
We comment that the above thermal ground state solution is not unique because $F(\vec{p})$ is arbitrary. Uniqueness could be accomplished by minimizing an operator representing free energy \cite{Umezawa:1975,Takahashi:1996zn} in general. But, for our case we found more practical way to determine $F(\vec{p})$ from two-point function. Note that the background field is equal to the (equal-time) two point function of the vector field with respect to the thermal vacuum $\left|0(\beta)\right>$. i.e. 
\begin{eqnarray}
\Psi_0((\vec{x},i),(\vec{y},j))=\left<\Psi((\vec{x},i),(\vec{y},j))\right>_\beta&=&\begin{pmatrix}
\left<\phi^a(\vec{x})\phi^a(\vec{y})\right>_\beta & i\left<\phi^a(\vec{x})\widetilde{\phi}^a(\vec{y})\right>_\beta\\
i\left<\widetilde{\phi}^a(\vec{x})\phi^a(\vec{y})\right>_\beta & -\left<\widetilde{\phi}^a(\vec{x}) \widetilde{\phi}^a(\vec{y})\right>_\beta\\
\end{pmatrix}
\end{eqnarray}
Using the Bogoliubov transformation (in \eqref{bogoliubov transformation1}), one can easily evaluate the vacuum expectation value $\left<\Psi(\vec{x}_1,\vec{x}_2;\vec{y}_1,\vec{y}_2)\right>_\beta$. Comparing to \eqref{background field}, one can determine $F(\vec{p})$ to be
\begin{equation}
F(\vec{p})=2\theta(\vec{p})=2\tanh^{-1} e^{-\beta|\vec{p}|}
\end{equation}

\section{ Collective modes and Bulk}
\label{sec:collective modes and bulk}

To study the collective modes one expands the bi-local field around thermal background\footnote{Here, we include numerical factor $i$ or $-1$ in the components of $\eta(t;(\vec{x},i),(\vec{y},j))$ in the same way as \eqref{collective field}. Moreover, we also include complex conjugate of the numerical factors in the components of $\pi(t;(\vec{x},i),(\vec{y},j))$ in order to keep canonical commutation relations.} :
\begin{eqnarray}
\Psi(t;(\vec{x},i),(\vec{y},j))&=&\Psi_0((\vec{x},i),(\vec{y},j))+{1\over \sqrt{N}} \eta(t;(\vec{x},i),(\vec{y},j))\label{large N expansion of bi-local field}\\
\Pi(t;(\vec{x},i),(\vec{y},j))&=&\sqrt{N}\pi(t;(\vec{x},i),(\vec{y},j))
\end{eqnarray}
By the Large $N$ expansion of the collective Hamiltonian, the quadratic Hamiltonian reads
\begin{equation}
H^{(2)}=2 \Tr\left[\pi\star \Psi_0\star \pi \right]+{1\over 8} \Tr\left[\Psi_0^{-1}\star \eta \star \Psi_0^{-1}\star \eta\star \Psi_0^{-1}\right]
\end{equation}
This quadratic Hamiltonian gives the equation of motion for $\eta$.
\begin{equation}
\ddot{\eta}=-{1\over 4}\eta\star \psi_0^{-1}\star \psi_0^{-1}-{1\over 2}\psi_0^{-1}\star\eta\star \psi_0^{-1}-{1\over 4}\psi_0^{-1}\star \psi_0^{-1}\star \eta\label{eom of fluctuation}
\end{equation}
To study the Hilbert space, it is convenient to express the quadratic Hamiltonian in the momentum space. 

By Fourier transformation of $\eta$ and $\pi$ into the momentums space where we use $e^{i\vec{p}_1\cdot \vec{x}-i\vec{p}_2\cdot \vec{y}}$ as kernel, the quadratic Hamiltonian becomes
\begin{equation}
H^{(2)}={1\over 2}\int d\vec{p}_1 d\vec{p}_2\left( \boldsymbol{\pi}^T(\vec{p}_1,\vec{p}_2) K(\vec{p}_1,\vec{p}_2)\boldsymbol{\pi}(\vec{p}_1,\vec{p}_2)+\boldsymbol{\eta}^T(\vec{p}_1,\vec{p}_2) V(\vec{p}_1,\vec{p}_2) \boldsymbol{\eta}(\vec{p}_1,\vec{p}_2)\right)
\end{equation}
where we define
\begin{equation}
\boldsymbol{\pi}(\vec{p}_1,\vec{p}_2)\equiv\begin{pmatrix}
\pi^{11}(\vec{p}_1,\vec{p}_2)\\
\pi^{12}(\vec{p}_1,\vec{p}_2)\\
\pi^{21}(\vec{p}_1,\vec{p}_2)\\
\pi^{22}(\vec{p}_1,\vec{p}_2)\\
\end{pmatrix}\quad,\quad
\boldsymbol{\eta}(\vec{p}_1,\vec{p}_2)\equiv\begin{pmatrix}
\eta^{11}(\vec{p}_1,\vec{p}_2)\\
\eta^{12}(\vec{p}_1,\vec{p}_2)\\
\eta^{21}(\vec{p}_1,\vec{p}_2)\\
\eta^{22}(\vec{p}_1,\vec{p}_2)\\
\end{pmatrix}
\end{equation}
\begin{eqnarray}
K(\vec{p}_1,\vec{p}_2)&\equiv&\begin{pmatrix}
c_1+c_2 & s_2 & s_1 & 0\\
s_2 & -c_1+c_2 & 0 & -s_1\\
s_1 & 0 & c_1-c_2 & -s_2\\
0 &  -s_1 &  -s_2 & -c_1-c_2\\
\end{pmatrix}\\
V(\vec{p}_1,\vec{p}_2)&\equiv& |\vec{p}_1|^2|\vec{p}_2|^2\begin{pmatrix}
c_1+c_2 & -s_2 & -s_1 & 0\\
-s_2 & -c_1+c_2 & 0 & s_1\\
-s_1 & 0 & c_1-c_2 & s_2\\
0 & s_1 &  s_2 & -c_1-c_2\\
\end{pmatrix}
\end{eqnarray}
and $c_i\equiv {\cosh(2\theta_i)\over |\vec{p}_i|}$, $s_i\equiv {\sinh(2\theta_i) \over |\vec{p}_i|}$ and $\theta_i\equiv \tanh^{-1} e^{-\beta|\vec{p}_i|}$ ($i=1,2$). To analyze the constraint structure, we Legendre transform to the Lagrangian scheme. To this end one has to express the momentum $\boldsymbol{\pi}$ in terms of $\dot{\boldsymbol{\eta}}$
\begin{equation}
\dot{\boldsymbol{\eta}}= K \boldsymbol{\pi}
\end{equation}
However, the matrix $K$ is not always invertible because the determinant of $K$ is given by
\begin{equation}
\det K={ (|\vec{p}_1|-|\vec{p}_2|)^2(|\vec{p}_1|+|\vec{p}_2|)^2\over |\vec{p}_1|^4|\vec{p}_2|^4}
\end{equation}
For $|\vec{p}_1|\ne |\vec{p}_2|$, the matrix $K$ is invertible so that Legendre transformation to Lagrangian is possible.
\begin{eqnarray}
\mathcal{L}&=&\boldsymbol{\pi}^T(\vec{p}_1,\vec{p}_2) \dot{\boldsymbol{\eta}}(\vec{p}_1,\vec{p}_2)- \mathcal{H}={1\over 2}\boldsymbol{\eta}^T\left(-K^{-1}\partial_t^2 -V\right)\boldsymbol{\eta}
\end{eqnarray}
By deviation, the equation of motion is given by
\begin{equation}
\left(-\partial_t^2 - KV\right) \boldsymbol{\eta}=0
\end{equation}
One can easily check that this agrees with \eqref{eom of fluctuation} as expected. 

On the other hand, for the case of $|\vec{p}_1|=|\vec{p}_2|$, the matrix $K$ is not invertible. We first diagonalize the matrix $K$
\begin{equation}
U^TKU=\mbox{diag}\left({2\sqrt{\cosh(4\theta)}\over |\vec{p}|},0,0,-{2\sqrt{\cosh(4\theta)}\over |\vec{p}|}\right)
\end{equation}
where $|\vec{p}|\equiv |\vec{p}_1|=|\vec{p}_2|$ and $\theta(|\vec{p}|)\equiv\theta_1(|\vec{p}_1|)=\theta_2(|\vec{p}_2|)$. The orthogonal matrix $U$ induces a canonical transformation from $(\boldsymbol{\eta}, \boldsymbol{\pi})$ to $(\overline{\boldsymbol{\eta}}, \overline{\boldsymbol{\pi}})$. i.e.
\begin{equation}
\overline{\boldsymbol{\pi}}\equiv U^T\boldsymbol{\pi}\quad,\quad \overline{\boldsymbol{\eta}}\equiv U^T\boldsymbol{\eta}
\end{equation}
Under this transformation, the quadratic Hamiltonian density for $|\vec{p}_1|=|\vec{p}_2|$ modes becomes
\begin{eqnarray}
{1\over 2}\left(\overline{\pi}^{11}{2\sqrt{\cosh(4\theta)}\over |\vec{p}|}\overline{\pi}^{11}-\overline{\pi}^{22}{2\sqrt{\cosh(4\theta)}\over |\vec{p}|}\overline{\pi}^{22}\right)+{1\over 2}\overline{\eta}^T W\overline{\eta}
\end{eqnarray}
where
\begin{equation}
W\equiv U^T VU=\begin{pmatrix}
a & b & 0 & 0 \\
b & 0 & 0 & b \\
0 & 0 & 0 & 0\\
0 & b & 0 & -a\\
\end{pmatrix}\quad\left(\mbox{with}\quad a\equiv{2 |\vec{p}|^3\over \sqrt{\cosh (4\theta)}}\;,\;b\equiv -{\sqrt{2}|\vec{p}|^3\sinh(4\theta)\over \sqrt{\cosh(4\theta)}}\right)
\end{equation}
Note that $\overline{\pi}^{12}$ and $\overline{\pi}^{21}$ do not appear in the Hamiltonian. In addition, the Hamiltonian contains a linear term in $\overline{\eta}^{12}$. Therefore, one can see that $\overline{\eta}^{12}(\vec{p}_1,\vec{p}_2)$ is a Lagrangian multiplier and the corresponding constraint $\mathcal{C}_1$ is given by
\begin{equation}
\mathcal{C}_1\equiv\overline{\eta}^{11}(\vec{p}_1,\vec{p}_2)+\overline{\eta}^{22}(\vec{p}_1,\vec{p}_2)=-\eta^{11}(\vec{p}_1,\vec{p}_2)+\eta^{22}(\vec{p}_1,\vec{p}_2)=0\qquad (\mbox{for}\quad |\vec{p}_1|=|\vec{p}_2|)
\end{equation}
In addition, $\left[H,\mathcal{C}_1\right]$ give a secondary constraint $\mathcal{C}_2$.
\begin{equation}
\mathcal{C}_2\equiv \overline{\pi}^{11}(\vec{p}_1,\vec{p}_2)-\overline{\pi}^{22}(\vec{p}_1,\vec{p}_2)=0\qquad (\mbox{for}\quad |\vec{p}_1|=|\vec{p}_2|)
\end{equation}
Note that $\mathcal{C}_1$ and $\mathcal{C}_2$ form the first class constraints. e.g. $\left[\mathcal{C}_1,\mathcal{C}_2\right]=0$. We further analyze these constraints in Appendix~\ref{sec:constraints}.

To express the fluctuations in terms of of creation and annihilation operators in the Large $N$, we need to solve the equation of motion \eqref{eom of fluctuation} in general. However, we do not have to explicitly solve \eqref{eom of fluctuation} because for the present problem we know the exact $O(N)$ singlet eigenstates in the Fock space. They are given in terms of the bi-local operators defined in Appendix~\ref{sec:algebra of bi-local operators}. For instance, consider $\Psi^{11}(t;\vec{x},\vec{y})$. The fluctuation $\eta^{11}$ can be obtained by subtracting the background field from $\Psi^{11}$, or equivalently, by normal ordering of $\Psi^{11}(t;\vec{x},\vec{y})$ with respect to the thermal vacuum $\left|0(\beta)\right>$. Recall that the annihilation operators $a^i(\vec{p})$ and $\widetilde{a}^i(\vec{p})$ do not annihilate the thermal vacuum, but the Bogoliubov transformed annihilation operators $a^i_\theta(\vec{p})$ and $\widetilde{a}^i_\theta(\vec{p})$ do. Therefore, for the correct normal ordering with respect to the thermal vacuum, we have to follow the standard normal ordering rule with $a^i_\theta(\vec{p}), a^{i\dag}_\theta(\vec{p}),\widetilde{a}^i_\theta(\vec{p})$ and $\widetilde{a}^{i\dag}_\theta(\vec{p})$. From either way (subtracting or normal-ordering), one can express $\eta^{11}(t;\vec{x},\vec{y})$ in terms of normal ordered $O(N)$ invariant composite operators of $a^i_\theta(\vec{p}), a^{i\dag}_\theta(\vec{p}),\widetilde{a}^i_\theta(\vec{p})$ and $\widetilde{a}^{i\dag}_\theta(\vec{p})$. We summarize the algebra of those bi-local composite operators in Appendix~\ref{sec:algebra of bi-local operators} together with a systematic $1/N$ expansion. Bi-local creation operators $\alpha^\dag,\widetilde{\alpha}^\dag, \gamma^\dag$ are defined to be the composite operators with the two creation operators:
\begin{eqnarray}
\begin{pmatrix}
\alpha^\dag(\vec{p}_1,\vec{p}_2) & \gamma^\dag(\vec{p}_1,\vec{p}_2) \\
\gamma^\dag(\vec{p}_2,\vec{p}_1) & \widetilde{\alpha}^\dag(\vec{p}_1,\vec{p}_2)
\end{pmatrix}\equiv{1\over \sqrt{2N |\vec{p}_1| |\vec{p}_2|} }\begin{pmatrix}
a_\theta^{i\dag}(\vec{p}_1)a_\theta^{i\dag}(\vec{p}_1) & a_\theta^{i\dag}(\vec{p}_1)\widetilde{a}_\theta^{i\dag}(\vec{p}_1)\\
\widetilde{a}_\theta^{i\dag}(\vec{p}_1)a_\theta^{i\dag}(\vec{p}_1) & \widetilde{a}_\theta^{i\dag}(\vec{p}_1)\widetilde{a}_\theta^{i\dag}(\vec{p}_1)\\
\end{pmatrix}\label{eq:bi-local operator1}
\end{eqnarray}
and one can also define their conjugate bi-local annihilation operators $\alpha(\vec{p}_,1,\vec{p}_2),\widetilde{\alpha}(\vec{p}_,1,\vec{p}_2), \gamma(\vec{p}_1,\vec{p}_2)$. Note that the physical creation operators are defined with respect to the thermal vacuum $|0(\beta)\rangle$. They satisfy the commutation relations:
\begin{eqnarray}
\left[\alpha(I,J),\alpha^\dag(K,L)\right]&=&{1\over 2}\left(\delta_{I,K}\delta_{J,L}+\delta_{I,L}\delta_{J,K}\right)\cr
\left[\widetilde{\alpha}(I,J),\widetilde{\alpha}^\dag(K,L)\right]&=&{1\over 2}\left(\delta_{I,K}\delta_{J,L}+\delta_{I,L}\delta_{J,K}\right)\cr
\left[\gamma(I,J),\gamma^\dag(K,L)\right]&=&{1\over 2}\delta_{I,K}\delta_{J,L}
\end{eqnarray}
where $I,J, \cdots$ represent the bi-local momentum such as $(\vec{p},\vec{q})$. One can express all $O(N)$ invariant operators in terms of these bi-local oscillators in $1/N$ expansion. For example, bi-local composite operators with two annihilation operators can be expanded as follows.
\begin{eqnarray}
{1\over \sqrt{2N |\vec{p}_1| |\vec{p}_2|}}\begin{pmatrix}
a_\theta^i(\vec{p}_1)a_\theta^i(\vec{p}_1) & a_\theta^i(\vec{p}_1)\widetilde{a}_\theta^i(\vec{p}_1)\\
\widetilde{a}_\theta^i(\vec{p}_1)a_\theta^i(\vec{p}_1) & \widetilde{a}_\theta^i(\vec{p}_1)\widetilde{a}_\theta^i(\vec{p}_1)\\
\end{pmatrix}=\begin{pmatrix}
\alpha(\vec{p}_1,\vec{p}_2) & \gamma(\vec{p}_1,\vec{p}_2) \\
\gamma(\vec{p}_2,\vec{p}_1) & \widetilde{\alpha}(\vec{p}_1,\vec{p}_2)
\end{pmatrix}+\mathcal{O}\left({1\over N}\right)\label{eq:bi-local operator2}
\end{eqnarray}
and the rest of composite operators such as $a^{i\dag} a^i$ are sub-leading in the ${1\over N}$ expansion so that we ignore them. Therefore, the linearized fluctuation $\eta^{11}$ can be expressed in terms of the bi-local oscillators $\alpha,\gamma $ (and their conjugates) in $1/N$ expansion, and we can read off the leading terms:
\begin{eqnarray}
&&\eta^{11}(t;\vec{x},\vec{y})={1\over  (2\pi)^2 }\int {d\vec{p}_1 d\vec{p}_2\over \sqrt{2 |\vec{p}_1||\vec{p}_2|}}\cr
&&\times\left[\alpha(\vec{p}_1,\vec{p}_2)\cosh\theta_1\cosh\theta_2 e^{-i(|\vec{p}_1|+|\vec{p}_2|)t+i\vec{p}_1\cdot \vec{x}+i\vec{p}_2\cdot \vec{y}}+\widetilde{\alpha}(\vec{p}_1,\vec{p}_2)\sinh\theta_1\sinh\theta_2 e^{i(|\vec{p}_1|+|\vec{p}_2|)t-i\vec{p}_1\cdot \vec{x}-i\vec{p}_2\cdot \vec{y}}\right.\cr
&&\qquad+\left.\gamma(\vec{p}_1,\vec{p}_2)\cosh\theta_1\sinh\theta_2 \left(e^{-i(|\vec{p}_1|-|\vec{p}_2|)t+i\vec{p}_1\cdot \vec{x}-i\vec{p}_2\cdot \vec{y}}+e^{-i(|\vec{p}_1|-|\vec{p}_2|)t+i\vec{p}_1\cdot \vec{y}-i\vec{p}_2\cdot \vec{x}}\right)\right]\cr
&&+h.c.+\mathcal{O}\left({1\over \sqrt{N}}\right)\label{eq:col normal mode1}
\end{eqnarray}
In the same way, one can also find the mode expansion of $\eta^{22}(t;\vec{x},\vec{y}), \eta^{12}(t;\vec{x},\vec{y})$ and $\eta^{21}(t;\vec{x},\vec{y})$.~e.g.
\begin{eqnarray}
&&\eta^{12}(t;\vec{x},\vec{y})={1\over  (2\pi)^2 }\int {d\vec{p}_1 d\vec{p}_2\over \sqrt{2 |\vec{p}_1||\vec{p}_2|}}\cr
&&\times\left[\gamma(\vec{p}_1,\vec{p}_2)\left(\cosh\theta_1\cosh\theta_2 e^{-i(|\vec{p}_1|-|\vec{p}_2|)t+i\vec{p}_1\cdot \vec{x}-i\vec{p}_2\cdot \vec{y}} + \sinh\theta_1\sinh\theta_2  e^{-i(|\vec{p}_1|-|\vec{p}_2|)t+i\vec{p}_1\cdot \vec{y}-i\vec{p}_2\cdot \vec{x}}   \right) \right.\cr
&& \left.+\alpha(\vec{p}_1,\vec{p}_2)\cosh\theta_1\sinh\theta_2 e^{-i(|\vec{p}_1|+|\vec{p}_2|)t+i\vec{p}_1\cdot \vec{x}+i\vec{p}_2\cdot \vec{y}}+\widetilde{\alpha}(\vec{p}_1,\vec{p}_2)\sinh\theta_1\cosh\theta_2 e^{i(|\vec{p}_1|+|\vec{p}_2|)t-i\vec{p}_1\cdot \vec{x}-i\vec{p}_2\cdot \vec{y}}\right]\cr
&&+h.c.+\mathcal{O}\left({1\over \sqrt{N}}\right)\label{eq:col normal mode2}
\end{eqnarray}
Using these mode expansions, one can evaluate the two-point Green's function in the Large $N$ limit. And, one can confirm that the Green's function satisfy the equation of motion \eqref{eom of fluctuation} (See Appendix~\ref{green's function}).

\subsection{Bulk Interpretation}
\label{sec:Bulk interpretation}

At zero temperature, the bi-local fields are mapped into the bulk fields of higher spin theory in the AdS background. This is simply accomplished in the light-cone gauge~\cite{Koch:2010cy,Mintun:2014gua} and was furthermore extended to the timelike (canonical) quantization in ~\cite{Koch:2014aqa}. At finite temperature, the TFD is expected to generate a space-time similar to the eternal black hole which asymptotically approaches AdS$_4$. The fluctuation modes found in the linearized collective analysis are expected to fully reproduce the bulk modes of higher spin in the modified space-time. Using the knowledge of the AdS/bi-local CFT map, we can work out their properties.

Let us summarize the main features of the canonical bi-local map to AdS bulk fields given in~\cite{Koch:2014aqa} for the $O(N)$ vector model CFT. We have the following expression for the bulk higher spin field 
\begin{eqnarray}
\mathcal{H}_{ s}^{(\pm)}(t;\vec{x},z)&=&C_{s,\pm}\int_{p^0>|\vec{p}|} \frac{d^2\vec{p}dp^0}{(2\pi)^3 2p^0} e^{-ip^0t+i\vec{p}\cdot\vec{x}}z^{\frac{1}{2}}\left[(p^0)^2-\vec{p}^2\right]^{-\frac{s}{2}+\frac{1}{4}}\cr
&&\qquad\qquad\times J_{\mp\frac{1}{2}} (\sqrt{(p^0)^2-\vec{p}^2}\; z) 
 \left[\mathcal{O}_s(p;\epsilon)\pm \mathcal{O}_s(p;\epsilon^*)\right]+\mbox{h.c}
\end{eqnarray}
where $(\pm)$ represents polarization of the bulk higher spin field. The spin-$s$ current $\mathcal{O}_s$ defining the Hilbert space of CFT is normal-ordered with respect to the vacuum (at zero temperature), which is the standard definition of CFT operator. Due to the standard normal-ordering, it can be expressed in term of bi-local operator~$\alpha(\vec{p}_1,\vec{p}_2)$:
%
\begin{equation}
\mathcal{O}^s(p;\epsilon)=\int_{\sin {\varphi_1-\varphi_2\over 2}>0} d\vec{p}_1 d\vec{p}_2\; \mathcal{J}(\vec{p}_1,\vec{p}_2)\delta^{(3)}(p_1^\mu+p_2^\mu-p^\mu)\left(\epsilon_\mu\cdot p_2^\mu\right)^{s} \frac{2^{3s+1}}{s!}\alpha(\vec{p}_1,\vec{p}_2)
\end{equation}
This map (and the kernel) directly follows from the following canonical relation between bi-local and AdS coordinates :
\begin{eqnarray}
\vec{p}&=&\vec{p}_1+\vec{p}_2\label{eq:map1}\\
p^z&=&2\sqrt{|\vec{p}_1| |\vec{p}_2|}\sin\left({\varphi_1-\varphi_2\over 2}\right)\label{eq:map2}\\
\theta&=&\arctan\left({2\vec{p}_2\times \vec{p}_1\over (|\vec{p}_2|-|\vec{p}_1|)p^z}\right)\label{eq:map3}
\end{eqnarray}
where $\vec{p}_i=(|\vec{p}_i|\cos\varphi_i,|\vec{p}_i|\sin\varphi_i)$. Under this map (with the conjugates determined by a chain rule) the bi-local on-shell condition is translates into an AdS on-shell condition:
\begin{equation}
(p^z)^2=(p^0)^2-\vec{p}^2\label{eq:on-shell condition}
\end{equation}
Here $p^z$ is a canonical conjugate to the radial AdS coordinate $z$. The center of momenta of the bi-local operator $\alpha(\vec{p}_1,\vec{p}_2)$, which is mapped to the (three) momentum $(p^0, \vec{p})$ of the spin-$s$ current operator, is always timelike at zero temperature. Hence, there is no bi-local operator with spacelike momentum at zero temperature, and the corresponding spin-$s$ current operator has no spacelike mode. Consequently, the constructed AdS bulk field propagates in $z$-direction ($p^z$ is real) at zero temperature.


For the thermal case, we suggest a similar reconstruction of the bulk field in terms of CFT operators. We will start with the discussion of the Hilbert space and give the explicit bulk reconstruction formula subsequently. Let us apply the bi-local canonical map given in the above to the finite temperature modes and their bulk interpretation. First we have the modes:
\begin{center}
{\setlength{\extrarowheight}{3pt}
\setlength{\tabcolsep}{15pt}
\begin{tabular}{|c|c|}
\hline
\multirow{2}{*}{$\alpha^\dag(\vec{p}_1,\vec{p}_2)$} & $p^0=|\vec{p}_1|+|\vec{p}_2|$ \\
 & $\vec{p}=\vec{p}_1+\vec{p}_2$\\
\hline
\multirow{2}{*}{$\widetilde{\alpha}^\dag(\vec{p}_1,\vec{p}_2)$} & $p^0=-|\vec{p}_1|-|\vec{p}_2|$\\
& $\vec{p}=-\vec{p}_1-\vec{p}_2$ \\
\hline
\end{tabular}
}
\end{center}
and, therefore, the operators $\alpha^\dag(\vec{p}_1,\vec{p}_2)$ and $\widetilde{\alpha}^\dag(\vec{p}_1,\vec{p}_2)$ create states in Hilbert space with timelike momentum. The map \eqref{eq:map1}$\sim$\eqref{eq:map2} and the on-shell condition \eqref{eq:on-shell condition} produces real values for the momentum $p^z$ in the bulk.
\begin{equation}
(p^z)^2=(p^0)^2-\vec{p}^2 \geqq 0
\end{equation}
On the other hand, our TFD collective analysis exhibits another set of modes corresponding~to:
\begin{center}
{\setlength{\extrarowheight}{3pt}
\setlength{\tabcolsep}{15pt}
\begin{tabular}{|c|c|}
\hline
\multirow{2}{*}{$\gamma^\dag(\vec{p}_1,\vec{p}_2)$} & $p^0=|\vec{p}_1|-|\vec{p}_2|$\\
& $\vec{p}=\vec{p}_1-\vec{p}_2$ \\
\hline
\end{tabular}
}
\end{center}
These are characterized by spacelike momentum, which leads to imaginary momentum $p^z$ of the bulk.
\begin{equation}
(p^z)^2=(p^0)^2-\vec{p}^2 \leqq 0
\end{equation}
We recognize the last set of modes as representing evanescent mode. In~\cite{Leichenauer:2013kaa,Rey:2014dpa}, the importance of such modes and their presence in black hole background asymptotic to AdS was discussed and emphasized. In particular their presence was seen to prevent a straightforward extension of the bulk to boundary kernel in the case of black hole type backgrounds.

Turning to our reconstruction of the bulk field in TFD formalism, we note the following. In recent work~\cite{Papadodimas:2012aq,Papadodimas:2013jku}, a construction was suggested based on the composite single trace operators, $\mathcal{O}$ and $\widetilde{\mathcal{O}}$ of CFT$_R$ and CFT$_L$, respectively, with the further argument that $\widetilde{\mathcal{O}}$ can be reconstructed from the Hilbert space of $\mathcal{O}$'s. 

In the bi-local field formalism, we have seen that the invariant Hilbert space is given in terms of three set of bi-local operators, $\alpha$'s, $\widetilde{\alpha}$'s and $\gamma$'s. The collective normal modes found in \eqref{eq:col normal mode1} and \eqref{eq:col normal mode2} are seen to be linear combinations of $\Psi^{11}\sim \phi^i\phi^i$, $\Psi^{12}\sim \phi^i\widetilde{\phi}^i$ and $\Psi^{22}\sim \widetilde{\phi}^i\widetilde{\phi}^i$.
 
The bi-local fields serve as generating fields for the spin-$s$ primary operators, and the reconstructed HS bulk fields can be then expressed in terms of these operators. This was explicitly given in our previous work at zero temperature. In the present TFD case, we are able to give an analogous reconstruction, however the operators will be seen to involve creation operators of the mixed Hilbert space described above. It is the additional set of mixed operators that is responsible for evanescent modes.
 
Let us consider the spin-$s$ conserved primary operators in the TFD of the free $O(N)$ vector model, given as :
\begin{eqnarray}
\mathcal{O}^{ab}_{s}(x;\zeta)\equiv \sum_{n=0}^{[s/2]}\sum_{k=0}^{s-2n}\frac{(-4)^n(-1)^k}{\sqrt{N}(2n)! k!(s-2n-k)!}:(\zeta\!\cdot\!\partial)^{n+k}\phi^i_{(a)}(x)\;(\zeta\!\cdot\!\partial)^{s-n-k}\phi^i_{(b)}(x) :
\end{eqnarray}
where $\phi^i_{(1)}(x)\equiv\phi^i(x)$ and $\phi^i_{(2)}(x)\equiv\widetilde{\phi}^i(x)$. Note that, for odd spin $s$, $\mathcal{O}^{12}_s$ and $\mathcal{O}^{21}_s$ are non-zero while $\mathcal{O}^{11}_s$ and $\mathcal{O}^{22}_s$ vanish automatically.

To find the representation of these operators on the invariant Hilbert space, we transform the currents into the momentum space. Any definition of CFT operators must involve a normal ordering procedure, and we emphasize that in our thermal case we define the spin-$s$ current operators through normal-ordering with respect to the thermal vacuum $|0(\beta)\rangle$. This has significant consequences. At zero temperature, the spin-$s$ current is consist of timelike bi-local operators due to a normal-ordering with respect to $|0\rangle=|0(\infty)\rangle$. On the other hand, the normal-ordering with respect to the thermal vacuum $|0(\beta)\rangle$ will imply that our spin-$s$ currents also have nonzero spacelike bi-local components. We have seen that $\alpha^\dag(\vec{p}_1,\vec{p}_2)$ and $\widetilde{\alpha}^\dag(\vec{p}_1,\vec{p}_2)$ create a state with the timelike momentum $(|\vec{p}_1|+|\vec{p}_2|,\vec{p}_1+\vec{p}_2)$ and $(-|\vec{p}_1|-|\vec{p}_2|,-\vec{p}_1-\vec{p}_2)$. On the other hand, $\gamma^\dag(\vec{p}_1,\vec{p}_2)$ create a state with the spacelike momentum $(|\vec{p}_>|-|\vec{p}_<|,\vec{p}_>-\vec{p}_<)$ where we define
\begin{equation}
\vec{p}_>=\begin{cases}
\vec{p}_1 & \mbox{when}\quad |\vec{p}_1|> |\vec{p}_2|\\
\vec{p}_2 & \mbox{when}\quad |\vec{p}_1|< |\vec{p}_2|\\
\end{cases}\quad,\quad \vec{p}_<=\begin{cases}
\vec{p}_2 & \mbox{when}\quad |\vec{p}_1|> |\vec{p}_2|\\
\vec{p}_1 & \mbox{when}\quad |\vec{p}_1|< |\vec{p}_2|\\
\end{cases}
\end{equation}
Hence, we have to separately Fourier transform $\mathcal{O}^{ab}_s(x;\zeta)$ for the timelike and the spacelike momentum denoted by $p^{\text{timeline}}$ and $p^{\text{spacelike}}$, respectively. After the Fourier transformation, as in~\cite{Koch:2014aqa}, we impose the conservation and traceless condition of the spin-$s$ primary operators with the timelike momentum to find physical operators. One can also repeat the same derivation for the case of the spacelike momentum. A solution of the two conditions is given by
\begin{equation}
\mathcal{O}^{ab}_{s}(p;\zeta)=\mathcal{O}^{ab}_{s}(p;\epsilon^*)\left(-\epsilon_\mu \zeta^\mu\right)^s+\mathcal{O}^{ab}_s(p;\epsilon)\left(-\epsilon^*_\mu \zeta^\mu\right)^s
\end{equation}
where a null polarization vector $\epsilon(p)$ for timelike momentum (and $\epsilon_\pm(p)$ for spacelike momentum) is given by
\begin{eqnarray}
\epsilon(p)&\equiv&\frac{1}{\sqrt{2}\left|\vec{p}\right|}\!\left(\frac{\vec{p}^2}{\sqrt{p^\mu p_\mu}},\frac{p^0p^1}{\sqrt{p^\mu p_\mu}}+ip^2,\frac{p^0p^2}{\sqrt{p^\mu p_\mu}}-i p^1\right) \qquad  (p\;:\mbox{timelike})\\
\epsilon_\pm(p)&\equiv&\frac{1}{\sqrt{2}\left|\vec{p}\right|}\!\left(\frac{i\vec{p}^2}{\sqrt{-p^\mu p_\mu}},\frac{ip^0p^1}{\sqrt{-p^\mu p_\mu}}\pm ip^2,\frac{ip^0p^2}{\sqrt{-p^\mu p_\mu}}\mp i p^1\right) \qquad (p\;:\mbox{spacelike})
\end{eqnarray}
Using them, we can then relate the physical spin-$s$ operators to the the bi-local oscillator fields. We find that, for $p^0>0$, 
\begin{eqnarray}
\mathcal{O}^{11}_{s}(p^{\text{timelike}};\epsilon)&=&{(-1)^s 2^{\frac{3}{2}s-2}\over (2\pi)^2 s!}   \int {d\vec{p}_1 d\vec{p}_2\over \sqrt{2|\vec{p}_1| |\vec{p}_2|}} \delta^{(3)}\left(p_1^\mu+p_2^\mu-p^\mu\right) \left(2\left|\vec{p}_1\right|\left|\vec{p}_2\right|-2\vec{p}_1\cdot \vec{p}_2\right)^{\frac{s}{2}} e^{is\Theta(\vec{p}_1,\vec{p}_2)}  \cr
&&\times  (\alpha(\vec{p}_1,\vec{p}_2)\cosh\theta_1\cosh\theta_2 +\widetilde{\alpha}^\dag(\vec{p}_1,\vec{p}_2)\sinh\theta_1\sinh\theta_2)\label{eq:timelike primary current11}\\
\mathcal{O}^{11}_{s}(p^{\text{spacelike}};\epsilon_+)&=&{(-i)^s 2^{\frac{3}{2}s-2}\over (2\pi)^2 s!}   \int {d\vec{p}_1 d\vec{p}_2\over \sqrt{2|\vec{p}_1| |\vec{p}_2|}}\; \delta^{(3)}\left(p_>^\mu-p_<^\mu-p^\mu\right) \left(2\left|\vec{p}_1\right|\left|\vec{p}_2\right|-2\vec{p}_1\cdot \vec{p}_2\right)^{\frac{s}{2}} e^{s\Phi(\vec{p}_1,\vec{p}_2)}  \cr
&&\times  (\gamma^\dag(\vec{p}_<,\vec{p}_>) \sinh\theta_>\cosh\theta_< +\gamma(\vec{p}_>,\vec{p}_<)\cosh\theta_>\sinh\theta_<)\label{eq:spacelike primary current11}
\end{eqnarray}
where $\Theta(\vec{p}_1,\vec{p}_2)$ and $\Phi(\vec{p}_1,\vec{p}_2)$ are functions defining the map between internal space coordinate $\theta$ of the higher spin field and the bi-local momentum space.
\begin{eqnarray}
\Theta(\vec{p}_1,\vec{p}_2)&\equiv& \arctan\left[ \frac{2\left(\vec{p}_2\times\vec{p}_1\right)}{\left(\left|\vec{p}_2\right|-\left|\vec{p}_1\right|\right)\sqrt{2\left|\vec{p}_1\right|\left|\vec{p}_2\right|-2\vec{p}_1\cdot \vec{p}_2}}\right]\\
\Phi(\vec{p}_1,\vec{p}_2)&\equiv& \tanh^{-1}\left[ \frac{2\left(\vec{p}_>\times\vec{p}_<\right)}{\left(\left|\vec{p}_>\right|+\left|\vec{p}_<\right|\right)\sqrt{2\left|\vec{p}_1\right|\left|\vec{p}_2\right|-2\vec{p}_1\cdot \vec{p}_2}}\right]
\end{eqnarray}
The physical operator with the $\epsilon^*$ polarization or complex conjugate of them have similar forms. For other spin-$s$ currents, we also have a similar result. At zero temperature ($\theta_{1,2}=0$), the spin-$s$ current $\mathcal{O}^{11}_{s}$ with spacelike momentum in \eqref{eq:spacelike primary current11} vanishes while the one with timelike momentum is expressed in term of only the bi-local operator $\alpha(\vec{p}_1,\vec{p}_2)$. This relation between spin-$s$ current operators and bi-local fields is a generalization to the thermal case of the AdS formulas of \cite{deMelloKoch:2012vc,Koch:2014aqa} .

For the completeness of the reconstruction, it will be important to work out the commutation relations of the above specified momentum space modes. We do that in Appendix~\ref{sec:commutators} where we show that after re-diagonalization (spin space) a canonical commuting set can be defined. Specifically from the relations (\ref{eq:commutation rel of timelike1}, \ref{eq:commutation rel of timelike2}, \ref{eq:commutation rel of spacelike1} and \ref{eq:commutation rel of spacelike2}), we establish the canonical operators $\mathcal{A}_s(k)$ (for $s\in \mathbb{Z}$) as:
\begin{eqnarray}
\mathcal{A}_s(k)\equiv
\sum_{s'=0}^\infty (\mathbb{K}^{-{1\over 2}}_{\beta,k})^{s,s'} \mathcal{O}^{11}_{s'}(k;\epsilon) + \sum_{s'=-1}^{-\infty} (\mathbb{K}^{-{1\over 2}}_{\beta,k})^{s,s'} \mathcal{O}^{11}_{|s'|}(k;\epsilon^*) \quad \left(\mbox{for} \quad k\;\mbox{: timelike}\right) \\
\mathcal{A}_s(k)\equiv \sum_{s'=0}^{\infty}(\mathbb{Q}^{-{1\over 2}}_{\beta,k})^{s,s'} \mathcal{O}^{11}_{s'}(k;\epsilon_+) + \sum_{s'=-1}^{-\infty} (\mathbb{Q}^{-{1\over 2}}_{\beta,k})^{s,s'} \mathcal{O}^{11}_{|s'|}(k;\epsilon_-)  \quad  (  \mbox{for} \quad k\;\mbox{: spacelike})
\end{eqnarray}
where we $\mathbb{K}_{\beta,k}$ is understood to be a matrix with indices $s$ and $s'$ so that $\mathbb{K}_{\beta,k}^{-{1\over 2}}$ is the square-root of the inverse of the matrix $\mathbb{K}_{\beta,k}$. One similarly constructs the canonical conjugates $\mathcal{A}^\dag_s(k)$ from $\mathcal{O}_s^{11\dag}$ in as well as a dual set $\widetilde{\mathcal{A}}_s(k)$ and $\widetilde{\mathcal{A}}^\dag_s(k)$ from $\mathcal{O}_s^{22}$ and $\mathcal{O}_s^{22\dag}$. Consequently, $\mathcal{A}_s$ and $\widetilde{\mathcal{A}}_s$ satisfy the canonical commutation relation.
\begin{gather}
\left[\mathcal{A}_s(k),\mathcal{A}^{\dag}_{s'}(q)\right]=\delta_{s,s'}\delta^{(3)}(k-q)\quad,\quad\left[\widetilde{\mathcal{A}}_s(k),\widetilde{\mathcal{A}}^{\dag}_{s'}(q)\right]=\delta_{s,s'}\delta^{(3)}(k-q)\cr
\left[\mathcal{A}_s(k),\widetilde{\mathcal{A}}_{s'}(q)\right]=\left[\mathcal{A}_s(k),\widetilde{\mathcal{A}}^{\dag}_{s'}(q)\right]=0
\end{gather}
and others vanish. It is central that all the commutators between the dual pairs vanish.

The vacuum expectation value of $\mathcal{A}^{\dag}_s\mathcal{A}_{s'}$ is
\begin{equation}
\left<\mathcal{A}^{\dag}_{s'}(k)\mathcal{A}_s(k)\right>={1\over e^{\beta k^0}-1}\delta_{s,s'}\delta^{(3)}(k-q) 
\end{equation}
and similar for $\left<\mathcal{A}_s(k)\mathcal{A}^{\dag}_{s'}(k)\right>$. Therefore, the vacuum expectation value of the number density operator of the spin-$s$ operator is
\begin{equation}
\left<\mathcal{A}^{\dag}_{s}(k)\mathcal{A}_s(k)\right>={1\over e^{\beta k^0}-1}\delta^{(3)}(k-q)
\end{equation}

We are now in a position to define the bulk higher spin fields $\mathcal{H}_s$ $(s\in\mathbb{Z})$ in terms of above constructed canonical operators $\mathcal{A}$'s and $\widetilde{\mathcal{A}}$'s. In the eternal black hole, one has four regions, \Romannum{1}$\sim$\Romannum{4} where region \Romannum{1} and region \Romannum{3} are right and left outside of the black hole and region \Romannum{2} (or, region \Romannum{4}) is inside of the black hole containing future (or, past) horizon, respectively (see the earlier work of~\cite{Herzog:2002pc,Kraus:2002iv}).

The presence of evanescent modes indicates that the effective potential $V(z)$ (in the language of~\cite{Rey:2014dpa}) is different from AdS case. To deduce the potential one would need the full bi-local map. This would define the wave functions $f_{s,k}(t,x,z)$ everywhere. The bulk field is given as follows

\begin{equation}
\mathcal{H}^{\text{\Romannum{1}}}_s(t,\vec{x},z)\equiv\int_{k^0>0} {dk^0 d\vec{p}\over (2\pi)^3} \left[f_{s,k}(t,\vec{x},z)\mathcal{A}_s(k)+f_{s,k}^*(t,\vec{x},z)\mathcal{A}^{\dag}_s(k)\right]
\end{equation}
where $f(t,\vec{x},z)$ is a canonically normalized wave function in the black hole background. In the region \Romannum{3}, one has
\begin{equation}
\mathcal{H}^{\text{\Romannum{3}}}_s(t,\vec{x},z)\equiv \int_{k^0>0} {dk^0 d\vec{p}\over (2\pi)^3}\left[ f_{s,k}^*(t,\vec{x},z)\widetilde{\mathcal{A}}_s(k)+f_{s,k}(t,\vec{x},z)\widetilde{\mathcal{A}}^{\dag}_s(k)\right]
\end{equation}
Here, we construct $\mathcal{H}^{\text{\Romannum{3}}}_s$ in such a way that $\widetilde{\mathcal{A}}^\dag_s(k)$ creates a mode $f_k(t,\vec{x},z)$ because the time direction in the region \Romannum{3} is opposite to that of the region \Romannum{1}. Moreover, the higher spin field inside of the black hole is
\begin{equation}
\mathcal{H}^{\text{\Romannum{2}}}_s(t,\vec{x},z)\equiv \int_{k^0>0} {dk^0 d\vec{p}\over (2\pi)^3}\left[ f^{(1)}_{s,k}(t,\vec{x},z)\mathcal{A}_s(k)+f^{(2)}_{s,k}(t,\vec{x},z)\widetilde{\mathcal{A}}^{\dag}_s(k)+h.c.\right]
\end{equation}
where $f^{(1)}_{s,k}(t,\vec{x},z)$ and $f^{(2)}_{s,k}(t,\vec{x},z)$ are two linearly independent solutions in the region \Romannum{2}.

To summarize we have studied the fluctuations of the bi-local (`single trace') operators in $O(N)$ vector model CFT at finite temperature in the Hamiltonian Thermo field dynamics formalism. The invariant spectrum corresponding to the bulk modes was evaluated and was seen to produce new modes (in comparison with the zero temperature AdS case). An exact $1/N$ expansion for the composite higher spin operators was performed, and evaluated in terms of the bi-local modes. This allows a complete reconstruction of the bulk for all high spin fields.

The main features of the reconstructed bulk fields at finite temperature are the following: at zero temperature (and at leading Large $N$) the spin $s$ fields were directly given in terms of the spin-$s$ composite operator of the CFT. At nonzero temperature which we have presented ,this is not the case, we have mixing. In order to obtain spin-$s$ bulk creation-annihilation modes with canonical properties we had a linear combination of all spin CFT operators. In this already we see a difference from the construction of~\cite{Papadodimas:2012aq,Papadodimas:2013jku} where the basic premise was that the vacuum (AdS) correspondence persist with only the doubling of decoupled CFT operators. 

Our construction is based on the `larger' Hilbert space  of three classes of Bogoliubov transformed bi-locals. {\it\small i.e.} $a_\theta^\dag\cdot~a_\theta^\dag$, $\widetilde{a}_\theta^\dag\cdot~\widetilde{a}_\theta^\dag$ and $a_\theta^\dag\cdot \widetilde{a}_\theta^\dag$. These bi-locals follow from the fact (discussed in the Section~\ref{sec:introduction}) that only the diagonal $O(N)$ subgroup of the full $O(N)\times O(N)$ is gauged. The second central feature of our the construction is that CFT operators are defined through normal ordering with respect to the mixed thermal vacuum state: $|0(\beta)\rangle$. As a consequence, the spin-$s$ current operators $\mathcal{O}_s^{11}\sim a^\dag\cdot~a^\dag$ and $\mathcal{O}_s^{22}\sim \widetilde{a}^\dag\cdot~\widetilde{a}^\dag$ of the Left and Right CFT are represented in terms of the above `enlarged' Hilbert space. Specifically, the first two classes of the bi-locals are packaged into timelike modes of $\mathcal{O}_s^{11}$ and $\mathcal{O}_s^{22}$ while the third one is packaged into spacelike modes of both $\mathcal{O}_s^{11}$ and $\mathcal{O}_s^{22}$. Especially, the third class of bi-locals are seen to have the features identical to `evanescent' modes present in black hole and other nontrivial backgrounds. These modes are crucial features of the bulk Higher spin fields that we construct.

We have established that the spin-$s$ current $\mathcal{O}_s^{11}$ on the right CFT commutes with the spin-$s'$ current $\mathcal{O}_s^{22}$ on the left CFT as expected. Moreover, it has been shown that the spacelike mode of the current, which encodes the evanescent mode, also commute with the timelike mode. However, since current operators with different spins do not commute, we have constructed the bulk fields from $\mathcal{A}_s$ and $\widetilde{\mathcal{A}}_s$ which diagonalize the commutation relation (in spin space).

\section{Conclusion}
\label{sec:conclusion}

We have studied in this paper the bi-local theory of the Large $N$ Thermo field CFT the $O(N)$ vector model and worked out the corresponding bulk Higher Spin realization. The bi-local composite operators in this case provide a simple set of `single-trace' operators, and their $1/N$ dynamics is faithfully given by the collective Hamiltonian. This was solved after the linearization around the (thermal) background with the modes and states explicitly constructed. They are seen to be related to the primary higher spin operators whose Large $N$ limit is evaluated and given in terms of collective modes. These represent the bulk fluctuating modes of the dual Higher spin theory. Several key features associated with the present construction are worth summarizing.

The bi-local field representation is seen to necessarily involve mixed operators in terms of Left and Right $O(N)$ theories. The larger class of operators is associated with the fact that we have found it appropriate to impose diagonal gauging of the $O(N)\times O(N)$ symmetry group. Furthermore, when defining primary currents and operators of the CFTs, a specification of the vacuum (and the process of normal ordering) is required. The vacuum chosen in the present construction is the nontrivial thermal state of \eqref{eq:single constraint}. The corresponding Bogoliubov transformation specifies physical creation-annihilation operators which again lead to mixed bi-locals. We have then shown the emergence of extra `evanescent' modes which were seen to be directly associated with the modes such as the mixed bi-locals between the Left and the Right CFT.

The extra modes are seen to play a significant role regarding the bulk Higher spin fields that we reconstruct. One could add that we can formally interpret the occurrence of spacelike modes of $\mathcal{O}_s^{11}$ and $\mathcal{O}_s^{22}$ as an analytic continuation\footnote{We thank Soo-Jong Rey for this interpretation} of timelike modes characteristic of zero-temperature AdS/CFT. At finite temperature, the analytic continuation enlarges the momentum space of operators to include the evanescent modes. In this paper, we gave explicit Hilbert space representation for such modes.

We have explicitly established the canonical commutation relations of the bulk modes that we have constructed. Finally the reconstructed bulk-fields involve mixing of all spins, and as such are more complex than the constructions given in earlier works.
 This construction, which we have demonstrated in the Large $N$ limit, suggests that the ideas of reconstructing the bulk as suggested in various proposals involving `subregion duality' likely do not hold. Among various possibilities, our construction might be most closely relate to the `wormhole' possibility of~\cite{Mathur:2014dia}. Some other implications on the reconstruction of bulk from bi-locals (precursors) were recently given in~\cite{Mintun:2015qda}. It is seen in the collective construction that decoupled CFT's do not necessarily lead to decoupled collective fields. This issue is seen to be closely related to the issue of gauging the $O(N)$ symmetry as to produce an enlarged singlet Hilbert space.

\acknowledgments

We are grateful to Samir Mathur and Soo-Jong Rey for helpful discussion. The work of AJ and JY is supported by the Department of Energy under contract DE-SC0010010.

\appendix

\section{Algebra of Bi-local Operators}
\label{sec:algebra of bi-local operators}

\subsection{Bi-local Operators}

In TFD of the $O(N)$ vector model, one can construct $O(N)$ invariant composite operators
\begin{eqnarray}
A_\theta(\vec{p}_1,\vec{p}_2)&\equiv& {1\over \sqrt{2N}}\sum_{i=1}^N {a_\theta^i(\vec{p}_1) \cdot a^i_\theta(\vec{p}_2)\over\sqrt{ |\vec{p}_1| |\vec{p}_2|}} \;,\quad \widetilde{A}_\theta(\vec{p}_1,\vec{p}_2)\equiv {1\over \sqrt{2N}}\sum_{i=1}^N {\widetilde{a}_\theta^i(\vec{p}_1) \cdot \widetilde{a}_\theta^i(\vec{p}_2)\over\sqrt{ |\vec{p}_1| |\vec{p}_2|}}\\
B_\theta(\vec{p}_1,\vec{p}_2)&\equiv&{1\over 2}\sum_{i=1}^N {a_\theta^{i\dag}(\vec{p}_1) \cdot a_\theta^i(\vec{p}_2)\over\sqrt{ |\vec{p}_1| |\vec{p}_2|}}\;,\quad \widetilde{B}_\theta(\vec{p}_1,\vec{p}_2)\equiv{1\over 2}\sum_{i=1}^N {\widetilde{a}_\theta^{i\dag}(\vec{p}_1) \cdot \widetilde{a}_\theta^i(\vec{p}_2)\over\sqrt{ |\vec{p}_1| |\vec{p}_2|}}\\
C_\theta(\vec{p}_1,\vec{p}_2)&\equiv& {1\over \sqrt{2N}}\sum_{i=1}^N {a^i_\theta(\vec{p}_1)\cdot \widetilde{a}^i_\theta(\vec{p}_2)\over\sqrt{ |\vec{p}_1| |\vec{p}_2|}} \;,\quad D_\theta(\vec{p}_1,\vec{p}_2)\equiv{1\over 2} \sum_{i=1}^N {a_\theta^i(\vec{p}_1) \cdot \widetilde{a}_\theta^{i\dag}(\vec{p}_2)\over\sqrt{ |\vec{p}_1| |\vec{p}_2|}}  
\end{eqnarray}
and their hermitian conjugates, $A^\dag_\theta(\vec{p}_1,\vec{p}_2),\widetilde{A}^\dag_\theta(\vec{p}_1,\vec{p}_2), C^\dag_\theta(\vec{p}_1,\vec{p}_2)$ and $D^\dag_\theta(\vec{p}_1,\vec{p}_2)$. Moreover, total Hamiltonian and momentum are defined by
\begin{eqnarray}
H_{\text{TFD}}&=&H-\widetilde{H}\;,\quad \mathbb{P}_{\text{TFD}}=\mathbb{P}-\widetilde{\mathbb{P}}
\end{eqnarray}
where
\begin{eqnarray}
H&=&=\sum_{\vec{p}}\omega_{\vec{p}} B_\theta(\vec{p},\vec{p})\;,\quad\mathbb{P}=\sum_{\vec{p}}\vec{p} B_\theta(\vec{p},\vec{p})\\
\widetilde{H}&=&\sum_{\vec{p}}\omega_{\vec{p}} B_\theta(\vec{p},\vec{p})\;,\quad\widetilde{\mathbb{P}}=\sum_{\vec{p}}\vec{p} \widetilde{B}_\theta(\vec{p},\vec{p})
\end{eqnarray}
To study algebra of these bi-local composite operators, it is convenient to define a matrix of bi-local composite operators. 
\begin{equation}
\mathbb{A}_\theta((\vec{p},i),(\vec{q},j))\equiv\begin{pmatrix}
A_\theta(\vec{p},\vec{q}) & C_\theta(\vec{p},\vec{q})\\
C_\theta(\vec{q},\vec{p}) & \widetilde{A}_\theta(\vec{p},\vec{q})\\
\end{pmatrix}\quad,\quad
\mathbb{A}_\theta^\dag((\vec{p},i),(\vec{q},j))\equiv\begin{pmatrix}
A_\theta^\dag(\vec{p},\vec{q}) & C_\theta^\dag(\vec{p},\vec{q})\\
C_\theta^\dag(\vec{q},\vec{p}) & \widetilde{A}_\theta^\dag(\vec{p},\vec{q})\\
\end{pmatrix}
\end{equation}
and
\begin{equation}
\mathbb{B}_\theta((\vec{p},i),(\vec{q},j))\equiv\begin{pmatrix}
B_\theta(\vec{p},\vec{q}) & D_\theta^\dag(\vec{p},\vec{q})\\
D_\theta(\vec{q},\vec{p}) & \widetilde{B}_\theta(\vec{p},\vec{q})\\
\end{pmatrix}
\end{equation}
Then, the commutation relation of $\mathbb{A}_\theta$ and $\mathbb{A}^\dag_\theta$ is given by
\begin{eqnarray}
&&\left[\mathbb{A}_\theta((I,i),(J,j)),\mathbb{A}_\theta^\dag((K,k),(L,l))\right]\cr
&=&{1\over 2} (\delta_{I,L}\delta_{i,l}\delta_{J,K}\delta_{j,k}+\delta_{I,K}\delta_{i,k}\delta_{J,L}\delta_{j,l})+{1\over N}\left(\delta_{I,K}\delta_{i,k}\mathbb{B}_\theta((J,j),(L,l))+\delta_{I,L}\delta_{i,l}\mathbb{B}_\theta((J,j),(K,k))\right)\cr
&&\qquad\qquad\qquad\qquad\qquad+{1\over N}\left(\delta_{J,K}\delta_{j,k}\mathbb{B}_\theta((J,j),(L,l))+\delta_{J,L}\delta_{j,l}\mathbb{B}_\theta((I,i),(K,k))\right)\label{eq:bi-local commutation relation1}
\end{eqnarray}
In Large $N$ limit, $\mathbb{A}_\theta$ and $\mathbb{A}_\theta^\dag$ are annihilation and creation operators, respectively. One can evaluate other commutation relations.
\begin{eqnarray}
\left[\mathbb{B}_\theta((I,i),(J,j)),\mathbb{A}_\theta^\dag((K,k),(L,l))\right]&=&{1\over 2} (\delta_{J,K}\delta_{j,k}\mathbb{A}_\theta^\dag((I,i),(L,l))+\delta_{J,L}\delta_{j,l}\mathbb{A}_\theta^\dag((K,k),(I,i)))\label{eq:bi-local commutation relation2}\\
\left[\mathbb{B}_\theta((I,i),(J,j)),\mathbb{A}_\theta((K,k),(L,l))\right]&=&-{1\over 2} (\delta_{I,K}\delta{i,k}\mathbb{A}_\theta^\dag((J,j),(L,l))+\delta_{I,L}\delta_{i,l}\mathbb{A}_\theta^\dag((K,k),(J,j)))\label{eq:bi-local commutation relation3}\\
\left[\mathbb{B}_\theta((I,i),(J,j)),\mathbb{B}_\theta((K,k),(L,l))\right]&=&{1\over 2}\left(\delta_{J,K}\delta_{j,k}\mathbb{B}_\theta((I,i),(L,l))-\delta_{I,L}\delta_{i,l}\mathbb{B}_\theta((K,k),(J,j))\right)\label{eq:bi-local commutation relation4}
\end{eqnarray}
Note that there are bi-local composite operators which commute with the total TFD Hamiltonian and the total TFD momentum:
\begin{eqnarray}
\left[H_{\text{TFD}},C_\theta(I,J)\right]=0\;,\quad \left[H_{\text{TFD}},C_\theta^\dag(I,J)\right]=0 \qquad \mbox{for}\quad |\vec{p}_I|=|\vec{p}_J|\\
\left[\mathbb{P}_{\text{TFD}},C_\theta(I,I)\right]=0\;,\quad \left[\mathbb{P}_{\text{TFD}},C_\theta^\dag(I,I)\right]=0 \qquad (\mbox{no summation})
\end{eqnarray}
We construct coherent states associated with $C_\theta(I,I)$'s.
\begin{equation}
\left|\{z(\vec{p)}\}\right>=\prod_{\vec{p}}\exp\left[z(\vec{p})C_\theta^\dag(\vec{p},\vec{p})-z^*(\vec{p}) C_\theta(\vec{p},\vec{p})\right]\left|0\right>
\end{equation}
Especially, the thermal vacuum is a specific coherent state with $z(\vec{p})= \theta(\vec{p})$. 

Going back to the bi-local composite operators, we want to study the Large $N$ expansion of the bi-local operators and express them in terms of canonical conjugate pair of operators. At zero temperature, the Large $N$ expansion of the bi-local operators was already found for the case of the $O(N)$ vector model~\cite{deMelloKoch:2012vc} and the $U(N)$ vector model~\cite{Koch:2014aqa} using Holstein-Primakoff transformation. One can repeat the exactly same procedure for the finite temperature $O(N)$ vector model because the only difference is that the bi-local momentum space of TFD is doubled. First, we consider the $O(N)$ invariant Fock space consisting of 
\begin{equation}
\mathbb{A}_\theta^\dag((I_1,i_1),(J_1,j_1))\cdots \mathbb{A}_\theta^\dag((I_n,i_n),(J_n,j_n))  \left|0(\theta)\right>
\end{equation}
In this singlet sector, a Casimir constraint is given by
\begin{eqnarray}
&&\mathbb{A}_\theta^\dag((I,i),(J,j))\star \mathbb{A}_\theta((J,j),(K,k)) - \mathbb{B}_\theta((I,i),(K,k))\cr
&&-{2\over N} \left(\mathbb{B}_\theta((I,i),(J,j))\star \mathbb{B}_\theta((J,j),(K,k))-{1\over 2}(\delta(0)+1) \mathbb{B}_\theta((I,i),(K,k))\right)=0\label{eq:Casimir constraint}
\end{eqnarray}
To solve the constraint \eqref{eq:Casimir constraint}, we introduce $\alpha^\dag((I,i),(J,j))$ given by
\begin{equation}
\alpha^\dag((I,i),(J,j))\equiv\begin{pmatrix}
\alpha^\dag(I,J) & \gamma^\dag(I,J)\\
\gamma^\dag(J,I) & \widetilde{\alpha}^\dag(I,J)\\
\end{pmatrix}\equiv \mathbb{A}_\theta^\dag((I,i),(J,j))
\end{equation}
and its conjugate $\alpha((I,i),(J,j))$ satisfying
\begin{equation}
\left[\alpha((I,i),(J,j)),\alpha^\dag((K,k),(L,l))\right]={1\over 2}(\delta_{I,K}\delta_{i,k}\delta_{J,L}\delta_{j,l}+\delta_{I,L}\delta_{i,l}\delta_{J,K}\delta_{j,k})
\end{equation}
Then, we present ansatz for $\mathbb{B}_\theta((I,i),(K,k))$.
\begin{eqnarray}
\mathbb{B}_\theta((I,i),(K,k))=\alpha^\dag((I,i),(J,j))\alpha((J,j),(K,k))
\end{eqnarray}
One can easily confirm that these ansatz satisfy \eqref{eq:bi-local commutation relation2} and \eqref{eq:bi-local commutation relation4}. Finally, one can solve the Casimir constraints for $\mathbb{A}_\theta((I,i),(J,j))$.
\begin{equation}
\mathbb{A}_\theta((I,i),(J,j))=\alpha((I,i),(J,j))+{2\over N}\alpha^\dag((K,k),(L,l))\alpha((L,l),(I,i))\alpha((J,j),(K,k))
\end{equation}
For our purpose, this realization in terms of $\alpha$'s is enough to see \eqref{eq:bi-local operator1} and \eqref{eq:bi-local operator2}. Also, we found other solutions for \eqref{eq:Casimir constraint} to get other realizations~\cite{deMelloKoch:2012vc,Koch:2014aqa} (And, one of them agrees with~\cite{Mintun:2014gua}). Those realizations give the same result because they satisfy \eqref{eq:bi-local operator1} and \eqref{eq:bi-local operator2} in Large $N$ expansion. Note that the bi-local composite operators $\mathbb{A}$ and $\mathbb{B}$ can be expressed in terms of the canonical pair of the bi-local field $\Psi$ and $\Pi$, which can be expanded in $1/N$~\cite{deMelloKoch:2012vc}.

\section{Constraints}
\label{sec:constraints}

In Section~\ref{sec:collective modes and bulk}, we found the first class constraints. We analyze these constraints using mode expansion of the bi-local fluctuation. For $|\vec{p}_1|=|\vec{p}_2|$, the mode expansion of the primary constraint is given by
\begin{eqnarray}
\eta^{11}(t,\vec{p}_1,\vec{p}_2)-\eta^{22}(t,\vec{p}_1,\vec{p}_2)&=&{1\over \sqrt{2 |\vec{p}_1| |\vec{p}_2| }}  \left[(\alpha(\vec{p}_1,-\vec{p}_2)-\widetilde{\alpha}^\dag(\vec{p}_1,-\vec{p}_2) )e^{-i(|\vec{p}_1|+|\vec{p}_2|)t}\right.\cr
&&\qquad\qquad \left.+(\alpha^\dag(-\vec{p}_1,\vec{p}_2)- \widetilde{\alpha}(-\vec{p}_1,\vec{p}_2)) e^{i(|\vec{p}_1|+|\vec{p}_2|)t}\right]
\end{eqnarray}
%
%
The commutation relation of $H_{\text{TFD}}$ and the primary constraint gives the secondary constraints. Then, we have two simple constraints
\begin{eqnarray}
\alpha(\vec{p}_1,\vec{p}_2)-\widetilde{\alpha}^\dag(\vec{p}_1,\vec{p}_2)=0\quad, \quad \alpha^\dag(\vec{p}_1,\vec{p}_2)- \widetilde{\alpha}(\vec{p}_1,\vec{p}_2)=0\qquad \mbox{for}\quad |\vec{p}_1|=|\vec{p}_2|
\end{eqnarray}
where they form the first class constraints. Thus, we impose two gauge conditions on their canonical conjugates.
\begin{eqnarray}
{1\over \sqrt{2}}\left(\alpha(\vec{p}_1,\vec{p}_2)-\widetilde{\alpha}^\dag(\vec{p}_1,\vec{p}_2)\right)=0\quad,\quad {1\over \sqrt{2}}\left(\alpha^\dag(\vec{p}_1,\vec{p}_2)+\widetilde{\alpha}(\vec{p}_1,\vec{p}_2)\right)=0\\
{1\over \sqrt{2}}\left(\alpha^\dag(\vec{p}_1,\vec{p}_2)-\widetilde{\alpha}(\vec{p}_1,\vec{p}_2)\right)=0\quad,\quad {1\over \sqrt{2}}\left(\alpha(\vec{p}_1,\vec{p}_2)+\widetilde{\alpha}^\dag(\vec{p}_1,\vec{p}_2)\right)=0
\end{eqnarray}
Therefore, a solution for these constraints is
\begin{eqnarray}
\alpha(\vec{p}_1,\vec{p}_2)=\alpha^\dag(\vec{p}_1,\vec{p}_2)=\widetilde{\alpha}(\vec{p}_1,\vec{p}_2)=\widetilde{\alpha}^\dag(\vec{p}_1,\vec{p}_2)=0\qquad \mbox{for}\quad |\vec{p}_1|=|\vec{p}_2|
\end{eqnarray}

\section{Green's Function}
\label{green's function}

Using the mode expansion of the bi-local fluctuation, we will calculate Green's functions and confirm that they are solutions of the equation of motions \eqref{eom of fluctuation}. For simplicity, we assume $t>0$ and denote the following Green's function by $\Delta_i$ $(i=1,2,3,4)$.
\begin{eqnarray}
\Delta_1\equiv\left<\eta^{11}(t,\vec{x}_1,\vec{x}_2)\eta^{11}(0,\vec{y}_1,\vec{y}_2)\right>\\
\Delta_2\equiv\left<\eta^{12}(t,\vec{x}_1,\vec{x}_2)\eta^{11}(0,\vec{y}_1,\vec{y}_2)\right>\\
\Delta_3\equiv\left<\eta^{21}(t,\vec{x}_1,\vec{x}_2)\eta^{11}(0,\vec{y}_1,\vec{y}_2)\right>\\
\Delta_4\equiv\left<\eta^{22}(t,\vec{x}_1,\vec{x}_2)\eta^{11}(0,\vec{y}_1,\vec{y}_2)\right>
\end{eqnarray}
Then, we obtain, for example, 
\begin{eqnarray}
&&\left<\eta^{11}(t,\vec{x}_1,\vec{x}_2)\eta^{11}(0,\vec{y}_1,\vec{y}_2)\right>\cr
&=&\int {d\vec{p}_1 d\vec{p}_2 \over (2\pi)^4 4|\vec{p}_1| |\vec{p}_2|}e^{i\vec{p}_1\cdot (\vec{x}_1-\vec{y}_1)+i\vec{p}_2\cdot (\vec{x}_2-\vec{y}_2)}\left( \cosh^2\theta_1e^{-i|\vec{p}_1| t}+\sinh^2\theta_1 e^{i|\vec{p}_1|t} \right)\left( \cosh^2\theta_2e^{-i|\vec{p}_2| t}+\sinh^2\theta_2 e^{i|\vec{p}_2|t} \right)    \cr
&&+(\vec{y}_1\;\longleftrightarrow\; \vec{y}_2)
\end{eqnarray}
One can confirm that they satisfy the equation of motions \eqref{eom of fluctuation}. Because the equation of motion is matrix equation, one have four differential equations related to $\Delta_i$'s. One of them is given by
\begin{eqnarray}
&&{\sqrt{\vec{\partial}_1^2 \vec{\partial}_2^2} \over \sqrt{-\vec{\partial}_1^2}+\sqrt{-\vec{\partial}_2^2}}\left[ (\partial_t^2-\vec{\partial}_1^2-\vec{\partial}_2^2)\Delta_1  \right.+2\sqrt{\vec{\partial}_1^2 \vec{\partial}_2^2} \cosh(2\theta_1)\cosh(2\theta_2)\Delta_1-2\sqrt{\vec{\partial}_1^2 \vec{\partial}_2^2}\cosh(2\theta_1)\sinh(2\theta_2)\Delta_2\cr
&&\qquad\qquad\left.-2\sqrt{\vec{\partial}_1^2 \vec{\partial}_2^2}\sinh(2\theta_1)\cosh(2\theta_2)\Delta_3+2\sqrt{\vec{\partial}_1^2 \vec{\partial}_2^2}\sinh(2\theta_1)\sinh(2\theta_2)\Delta_4 \right]\cr
&=&-{i \over 2} \cosh(\theta_1 +\theta_2)\cosh(\theta_1-\theta_2)\delta(t)\delta^{(2)}(\vec{x}_1-\vec{y}_1)\delta^{(2)}(\vec{x}_2-\vec{y}_2)+(\vec{y}_1\;\longleftrightarrow \;\vec{y}_2)
\end{eqnarray}
We also calculate the two-point function of $\mathcal{O}$'s in the momentum space.
\begin{eqnarray}
\left<\mathcal{O}_s^{11}(k;\epsilon)\mathcal{O}_{s'}^{11\dag}(q;\epsilon^*)\right>&=&{e^{\beta k^0}\over e^{\beta k^0}-1} \mathbb{K}_{\beta,k}^{s,s'}\delta^{(3)}(k^\mu-q^\mu)\\
\left<\mathcal{O}_{s'}^{11\dag}(q;\epsilon^*)\mathcal{O}_s^{11}(k;\epsilon)\right>&=&{1 \over e^{\beta k^0}-1} \mathbb{K}_{\beta,k}^{s,s'}\delta^{(3)}(k^\mu-q^\mu)
\end{eqnarray}
and similar for others.

\section{Commutators}
\label{sec:commutators}

It is useful to consider changes of variables from the bi-local momentum space $(\vec{p}_1,\vec{p}_2)$ to the three dimensional timelike or spacelike momentum space $p^\mu$ and coordinate of the internal space $\theta$ or $\phi$, respectively. These two changes of variables are induced from \eqref{eq:timelike primary current11} and \eqref{eq:spacelike primary current11}. First of all, a transformation to the timelike momentum and the internal coordinate space is given by
\begin{eqnarray}
p^0&=&|\vec{p}_1|+|\vec{p}_2|\\
\vec{p}&=&\vec{p}_1+\vec{p}_2\\
\theta&=&\Theta(\vec{p}_1,\vec{p}_2)
\end{eqnarray}
where the Jacobian of this transformation is given by
\begin{equation}
\mathcal{J}(\vec{p}_1,\vec{p}_2)\equiv {\sqrt{2|\vec{p}_1| |\vec{p}_2| -2\vec{p}_1\cdot \vec{p}_2} \over |\vec{p}_1| |\vec{p}_2|}
\end{equation}
Hence, an integration over the bi-local momentum space can be transformed as follows.
\begin{equation}
\int {d\vec{p}_1d\vec{p}_2\over |\vec{p}_1| |\vec{p}_2|} =\int_{p^0>|\vec{p}|} {d^3pd\over \sqrt{(p^0)^2-\vec{p}^2}}\theta
\end{equation}
Note that this change of variables is closely related to \eqref{eq:map1}$\sim$\eqref{eq:map3} for the case of AdS/CFT. In addition, a transformation to space like momentum and internal coordinate space is given by
\begin{eqnarray}
p^0&=&|\vec{p}_>|-|\vec{p}_<|\\
\vec{p}&=&\vec{p}_>-\vec{p}_<\\
\varphi&=&\Phi(\vec{p}_>,\vec{p}_<)
\end{eqnarray}
where the Jacobian of this transformation is also $\mathcal{J}(\vec{p}_1,\vec{p}_2)$. Therefore, an integration over the bi-local momentum space can also be changed into an integration over the spacelike momentum and the internal coordinate space.
\begin{equation}
\int d\vec{p}_1d\vec{p}_2 \mathcal{J}(\vec{p}_1,\vec{p}_2)=\int_{p^0<|\vec{p}|} d^3p \int d\varphi
\end{equation}

First of all, the commutation relations of $\mathcal{O}^{11}_s$ with the timelike momentum are
\begin{eqnarray}
\left[\mathcal{O}^{11}_s(k;\epsilon),\mathcal{O}^{11\dag}_{s'}(q;\epsilon)\right]&=&\mathbb{K}_{\beta,k}^{s,-s'}\delta^{(3)}(k^\mu-q^\mu)\qquad\mbox{for}\quad (k,q\;\mbox{: timelike})\label{eq:commutation rel of timelike1}\\
\left[\mathcal{O}^{11}_s(k;\epsilon),\mathcal{O}^{11\dag}_{s'}(q;\epsilon^*)\right]&=&\mathbb{K}_{\beta,k}^{s,s'}\delta^{(3)}(k^\mu-q^\mu)\qquad\mbox{for}\quad (k,q\;\mbox{: timelike})\label{eq:commutation rel of timelike2}
\end{eqnarray}
where a function $\mathbb{K}_{\beta,k}^{s,s'}$ is defined to be
\begin{eqnarray}
\mathbb{K}_{\beta,k}^{s,s'}\!&\equiv&\! \left(  (p^0)^2-\vec{p}^2 \right)^{{s+s' \over 2}-{1\over 2} }  I^{s,s'}_\beta(k)\\
I^{s,s'}_\beta(k)\!&\equiv& \!\frac{2^{\frac{3}{2}(|s|+|s'|)-6}}{ (2\pi)^4|s|!|s'|!}   \sinh{\beta k^0\over 2} \!\int \!  {d\theta \;\;e^{i(s-s')\theta} \over \sinh\!\left({\beta\over 4}(k^0-|\vec{k}|\cos\theta)\right)\sinh\!\left({\beta\over 4}(k^0+|\vec{k}|\cos\theta)\right)  }
\end{eqnarray}
Note that, at zero temperature, the function $\mathbb{K}_{\beta,k}^{s,s'}$ becomes
\begin{equation}
\lim_{\beta\rightarrow \infty} \mathbb{K}_{\beta,k}^{s,s'}\sim \left(  (k^0)^2-\vec{k}^2 \right)^{s-{1\over 2} } \delta_{s,s'}
\end{equation}
Hence, $\mathcal{O}^{11}_s(p;\epsilon)$ commutes with $\mathcal{O}^{11\dag}_s(p;\epsilon)$ at zero temperature, and is conjugate to $\mathcal{O}^{11\dag}_s(p;\epsilon^*)$ as expected.
In the similar way, one can calculate the commutation relations of $\mathcal{O}^{11}_s$'s with spacelike momentum given by
\begin{eqnarray}
\left[\mathcal{O}^{11}_s(k;\epsilon_+),\mathcal{O}^{11\dag}_{s'}(q;\epsilon_+)\right]&=&\mathbb{Q}^{s,-s'}_{\beta,k}\delta^{(3)}(k^\mu-q^\mu)\qquad\mbox{for}\quad (k,q\;\mbox{: spacelike})\label{eq:commutation rel of spacelike1}\\
\left[\mathcal{O}^{11}_s(k;\epsilon_+),\mathcal{O}^{11\dag}_{s'}(q;\epsilon_-)\right]&=&\mathbb{Q}^{s,s'}_{\beta,k}\delta^{(3)}(k^\mu-q^\mu)\qquad\mbox{for}\quad (k,q\;\mbox{: spacelike})\label{eq:commutation rel of spacelike2}
\end{eqnarray}
where a function $Q^{s,s'}_{\beta,k}$ are defined to be
\begin{eqnarray}
\mathbb{Q}^{s,s'}_{\beta,k}\!&\equiv&\! \left(  (p^0)^2-\vec{p}^2 \right)^{{s+s' \over 2}-{1\over 2} }  J^{s,s'}_\beta(k)\\
J^{s,s'}_\beta\!(k)\!&\equiv&\! \frac{i2^{\frac{3}{2}(|s|+|s'|)-6}}{(2\pi)^4 |s|!|s'|!}   \sinh{\beta k^0\over 2} \!\! \int \! \! { d\varphi\;\; e^{(s-s')\varphi} \over \sinh\left({\beta\over 4}(-k^0+|\vec{k}|\cosh\varphi)\right)\sinh\left({\beta\over 4}(k^0+|\vec{k}|\cosh\varphi)\right)  }
\end{eqnarray}
Contrast to $\mathbb{K}^{s,s'}_{\beta,k}$, the function $\mathbb{Q}^{s,s'}_{\beta,k}$ vanish at zero temperature, which is consistent with the fact that $\mathcal{O}^{11}$ with the spacelike momentum becomes zero at zero temperature.

It is easy to see that spin-$s$ operators with the timelike momentum commutes with spin-$s$ operator with the spacelike momentum. In addition, $\mathcal{O}^{11}$'s commute with $\mathcal{O}^{22}$'s. i.e.
\begin{equation}
\left[\mathcal{O}^{11}_s(k^{\text{timelike}}),\mathcal{O}^{11}_{s'}(q^{\text{spacelike}})\right]=\left[\mathcal{O}^{11}_s,\mathcal{O}^{22}_{s'}\right]=0
\end{equation}

\end{document}